\documentclass[%
 reprint,
 amsmath,amssymb,
 aps,
]{revtex4-2}
\usepackage{physics}
\usepackage{float}

\usepackage[pdftex, pdftitle={Article}, pdfauthor={Author}]{hyperref} % For hyperlinks in the PDF

\begin{document}

\title{Probing Clifford Algebras Through Spin Groups: A Standard Model Perspective}
    \author{Armando Reynoso}
    \email[Correspondence email address: ]{armando.reynoso01@student.csulb.edu}% Your name
    \affiliation{Department of Physics and Astronomy, California State University Long Beach, Long Beach, California 90840, USA}
   
\begin{abstract}
Division algebras have demonstrated their utility in studying non-associative algebras and their connection to the Standard Model through complex Clifford algebras. This article focuses on exploring the connection between these complex Clifford algebras and their corresponding real Clifford algebras providing insight into geometric properties of bivector gauge symmetries. We first generate gauge symmetries in the complex Clifford algebra through a general Witt decomposition. Gauge symmetries act as a constraint on the underlying real Clifford algebra, where they're then translated from their complex form to their bivector counterpart. Spin group arguments allow the identification of bivector structures which preserve the gauge symmetry yielding the corresponding real Clifford algebra. We conclude that Standard Model gauge groups emerge from higher-dimensional Clifford algebras carrying Euclidean signatures, where particle states are recognized as a combination of basis elements corresponding to complex Euclidean Clifford algebras.
\end{abstract}

\maketitle

\section{Introduction} \label{sec:outline}
Despite the success of the Standard Model in its predictive power in explaining interactions between matter and the three fundamental forces, understanding its group theoretical origin remains an active study. Unification models such as the Georgi-Glashow model, engulf the Standard Model gauge group into a larger single simple gauge group $SU(5)$, where symmetry breaking into the Standard Model occurs at some grand unified scale\cite{Georgi:1974sy}. Similarly, intricacies of the Standard Model have been explored not through unification, but rather through the exploration of non-associative algebras as first showed in\cite{cit6,cit7} by connecting split octonions to quark structure $SU(3)_{C}$.
\par
Further exploration of octonions and their role in the Standard Model led to the Dixon construction\cite{cit4}, expressing fermionic degrees of freedom as tensor product of division algebras. Division algebras further paved the study of non-associative algebras in the Standard Model by realizing a generation of particles in terms of $\mathbb{C}\otimes\mathbb{O}$ and its left adjoint algebra $(\mathbb{C}\otimes\mathbb{O})_{L}$ decomposed into minimal left ideals\cite{cit14}. Basis states from these ideals are then placed in correspondence with particle states identified by how they transform under unitary symmetries constructed from the automorphism group of the octonions $Aut(\mathbb{O})=G_{2}$\cite{cit14}. 
\par
The observation that one copy of octonions describes one generation of particles, a reasonable possibility in requiring three copies of octonions to described three generations has led towards the consideration of the exceptional Jordan algebra $J_{3}(\mathbb{O})$\cite{cittt3,cit2,cit3,cit888} playing a role in the Standard Model. Sedenions $\mathbb{S}$, the next division algebra after octonions generated form the Caley-Dickson construct has also been explored as a mathematical structure to represent three generation of particles\cite{cit12,cit13}. Further studies implementing algebras such as the Poincare group, braid group or $E_{6}$ and their relation to non-associative algebras describing aspects of the Standard Model have also been explored in the past\cite{cit1,cit8,cit10,cit11,citnew1}.
\par
A consequence on studying non-associative algebras leads to the construct of complex Clifford algebras such as $(\mathbb{C}\otimes\mathbb{O})_{L}\simeq\mathbb{C}l(6)$ in describing one generation of particles. Similarly $\mathbb{C}l(4)$ has been studied in its ability to describe weak interactions\cite{cit15,cit16} giving one family of leptons. Given that the complexification of Clifford algebras $\mathbb{C}\otimes Cl(p,q)=\mathbb{C}l(n)$ hides the underlying signature, it is worth exploring whether there exist a method to identify the appropriate indefinite signature, by looking at its underlying real bivector structure. This method is explored by first studying the connection between a complex Clifford algebra and its real Clifford Algebra through a general Witt decomposition\cite{witt1,witt2}. The Witt basis provides maximally isotropic sub-spaces forming a basis of our complex algebra, allowing us to construct gauge symmetries independent of the automorphism of non-associative algebras. Gauge symmetries are then mapped into their bivector form. The richness of real Clifford algebras structures\cite{Cliffbook1,Cliffbook3,Cliffbook2} in addition to the $Spin$ groups\cite{witt2} serving as a constraint allows us to identify indefinite signatures which describe Standard Model features prior to complexifying, yielding a deeper understanding between $Spin$ groups and complex Clifford algebras.
\par
In an age where the Standard Model carries itself as the predominant theory of fundamental interactions, one might consider studying the Standard Model purely through Clifford algebras a redundant reformulation of what is already known about nature. However, the counter argument is that Clifford algebras emphasizes the Standard Model and space-time structure from a geometric standpoint\cite{1cit10,1cit12,1cit13,1cit14,1cit15,1cit17,1cit19,1cit20,1cit21,1cit_1,1cit_2}. Clifford algebras provide the structure and versatility to probe ties between $Spin$ groups and their relation to the Standard Model, leading to a deeper grasp pertaining to the intimate constructs relating geometry, symmetries and fundamental interactions. It will provide the pathway to probe a complex Clifford spinor action as those explored in\cite{1cit_3,1cit_4}, and having a foundational understanding of how one might translate the theory into its bivector counterpart applying field theoretical techniques appropriate to Clifford algebras\cite{1cit9,1cit11,1cit16}, in addition to studying the connection between gauge groups and complex algebras. 
\par
The article is organized as follow: In Sec.\ref{sec:clifford} we review the defining properties of a Clifford algebra. In Sec.\ref{sec:complex} we develop the formalism where complexification allows us to study properties of real Clifford algebras through a generalized Witt decomposition accounting indefinite signatures. In Sec.\ref{sec:unitary} we employ the general Witt decomposition to arrive at the unitary symmetries, utilizing the $Spin$ group to identify the underlying Clifford algebra admitting the gauge group of our interest to explore its particle spectrum. We conclude the article and give outlooks in Sec\ref{sec:conclusion}.
\section{Clifford Algebras} \label{sec:clifford}
In this section, we review foundational aspects of Clifford algebras, providing a self-contained overview of key mathematical definitions\cite{Cliffbook1,Cliffbook2,Cliffbook3}. A pivotal point in the study of the Standard Model involves the introduction of Dirac matrices $\{\gamma_{0},\gamma_{1},\gamma_{2},\gamma_{3}\}$ satisfying the anti-commutation relation 
\begin{equation}\label{1eqn}
\{\gamma^{\mu},\gamma^{\nu}\}=\gamma^{\mu}\gamma^{\nu}+\gamma^{\nu}\gamma^{\mu}=2\eta^{\mu\nu}I_{4}
\end{equation}
To establish Lorentz invariance, we construct Spin $\frac{1}{2}$ representations of $SO^{+}(1,3)$ by forming generators of the form $S^{\mu\nu}=\frac{i}{4}[\gamma^{\mu},\gamma^{\nu}]$ satisfying the Lorentz commutation relation
\begin{equation}\label{2eqn}
    [S^{\mu\nu},S^{\rho\sigma}]=i(g^{\nu\rho}S^{\mu\sigma}-g^{\mu\rho}S^{\nu\sigma}-g^{\nu\sigma}S^{\mu\rho}+g^{\mu\sigma}S^{\nu\rho})
\end{equation}
Full reducibility of boost and rotation spinor representations implies that the Dirac representation of the Lorentz group is reducible, leading to the Weyl representation\cite{2cit1,Peskin:1995ev}. While the Dirac and Weyl representation differ from a unitary transformation, they're equivalent as a matrix classification of a complex Clifford algebra. That is to say, there exist a framework in which the algebra Eq.(\ref{1eqn}) can be encoded in a vector space independent of a representation. Mathematically a Clifford algebra is generated by initially defining a vector space $V$ equipped with a quadratic form $\phi(v)$ over a field $k$. The algebra is generated as the quotient space
\begin{equation}\label{Cliff}
    Cl(V,\phi)=\frac{\mathcal{T}(V)}{\mathcal{I}_{\phi}(V)}
\end{equation}
defining an equivalence relation with $\mathcal{T}(V)=\sum\otimes^{r}V$ as the tensor product of vector spaces and the ideal $\mathcal{I}_{\phi}(V)=v\otimes v-\phi(v)$, a special subset closed under the operation of multiplication by any element of the algebra. If $\{e_{i}:i=1,2,...,n\}$ is an orthonormal basis for $V$ with respect to $\phi(e_{i},e_{j})=\delta_{ij}$, then $Cl(V,Q)$ is generated by basis elements satisfying
\begin{equation}\label{3eqn}
    e_{i}e_{j}+e_{j}e_{i}=2\phi(e_{i},e_{j})
\end{equation}
A Clifford algebra Eq.(\ref{Cliff}) is spanned by $dimCl(p,q)=2^{p+q}$ possible unordered combinations composed of orthonormal basis vectors $\{e_{i}:i=1,2,...,n\}$ satisfying Eq.(\ref{3eqn}). For an indefinite signature $Cl(p,q)$, $p$ denotes the numbers of positive norm basis vectors, and $q$ the number of negative norm basis vectors. As an example, consider $Cl(3,0)$ generated by the basis vectors $\{e_{1},e_{2},e_{3}\}$. $Cl(3,0)$ is then spanned by $2^{3}$ unordered combinations given by
\begin{align}\label{5eqn}
    1&&\{e_{1},e_{2},e_{3}\}&&\{e_{1}e_{2}, e_{1}e_{3},e_{2}e_{3}\}&&\{e_{1}e_{2}e_{3}\}
\end{align}
Any element of the Clifford algebra has a definite grade $k$ with distinct basis elements $e_{i},...,e_{k}$. Grading as a mathematical tool provides both clarity and organization to our elements by providing a clear geometric/intuitive interpretation of elements. For example, grade 1 elements denotes basis vectors, grade 2 elements are bivector oriented plane segments, grade 3 elements representing oriented volumes, where we can represent $Cl(p,q)$ as a direct sum of graded sub-spaces
\begin{equation}
Cl(p,q)=W_{0}\oplus W_{1}\oplus...\oplus W_{p+q}
\end{equation}
The highest grade element is recognized as the pseudo-scalar of our algebra which squares to $I^{2}=-1$. 
The pseudo-scalar is valuable as it allows us to specify the vector space over which the algebra is defined. A commonly recognized isomorphism arises from the multiplication of a vector with the pseudo-scalar $Ie_{j}=(e_{1}e_{2}e_{3})e_{j}$ which evaluates to 
\begin{align}
    Ie_{3}=e_{1}e_{2}&&Ie_{1}=e_{2}e_{3}&&Ie_{2}=e_{3}e_{1}
\end{align}
identifying our relations with those of the Pauli algebra. These expressions highlight a key advantage of utilizing Geometric algebra, enabling representation-free approaches. However it's worth noting that through the classification of Clifford algebras, basis vectors can be mapped to matrix representations such as the case for the Dirac Clifford algerba \cite{2cit1}$\gamma^{0}\rightarrow\sigma^{3}\otimes I_{2}$ and $\gamma^{j}\rightarrow i\gamma^{2}\otimes\sigma^{j}$
satisfying Eq.(\ref{1eqn}) In this article we will be interested in bivector elements generating the exceptional isomorphism of the $Spin$ group which we explore in the following section.
\section{Complexification} \label{sec:complex}
In Sec.(\ref{sec:outline}), it was referenced that a Witt decomposition of left adjoint actions of $\mathbb{C}\otimes\mathbb{O}$ acting on itself generates $\mathbb{C}l(6)$ with basis states of ideals in $\mathbb{C}l(6)$ transforming as leptons and quarks under the unbroken symmetries $U(1)_{em}\times SU(3)_{C}$. These unitary symmetries follow from the observation that that $Aut(\mathbb{O})=G_{2}$ of octonions contains the strong interaction subgroup $SU(3)$ leading to the possibility of non-associative algebras and their role in the Standard Model\cite{cit14,cit15,cit16}. It is suggestive to find a general Witt decomposition from which we can generate complex algebras from various indefinite signatures. From our Witt decomposition we can then translate the operator unitary symmetries to the corresponding bivector algebras. A geometric approach to achieve complexification of a real Clifford algebra was alluded in the study of $Spin$ group\cite{witt2}. We start with a vector space 
\begin{equation}\label{2eqn1}
    \mathcal{R}^{n,n}=V_{e}^{n}\otimes V_{\bar{\phi}}^{*n}
\end{equation}
generated by an $n$-dimensional Euclidean vector space $V_{e}$ and its $n$-dimensional dual anti-Euclidean vector space $V^{*}_{\bar{\phi}_{i}}$. Our basis vectors can be expressed in terms of null vectors\cite{witt2} $\{w,w^{*}\}$
\begin{align}\label{2eqn2}
    e_{i}=\omega_{i}+\omega^{*}_{i}&&\bar{\phi}_{i}=\omega_{i}-\omega^{*}_{i}
\end{align}
where Eq.(\ref{2eqn1}) can be decomposed into $\mathcal{R}^{n,n}=\mathcal{R}_{e}^{n,0}\otimes\mathcal{R}_{\bar{\phi}_{i}}^{0,n}$implying $\omega_{i}\omega^{*}_{j}+\omega^{*}_{j}\omega_{i}=\delta_{ij}$
resembling the fermionic creation and annihilation operator relation with
\begin{align}\label{2eqn22}
    \omega_{i}=\frac{1}{2}(e_{i}+\bar{\phi}_{i})&&\omega^{*}=\frac{1}{2}(e_{i}-\bar{\phi}_{i})
\end{align}
recasting our variables moving forward as $\omega_{i}\rightarrow a_{i}$ and $\omega^{*}_{i}\rightarrow a^{\dagger}_{i}$. In laying the basis for $\mathbb{C}l(n)$, we establish a complex structure by postulating the existence of a complex unit scalar $i$. The complex structure allows a Clifford algebra e.g. $Cl(4,0)$ with orthonormal basis $\{e_{1},e_{2},\phi_{1},\phi_{2}\}$to satisfy Eq.(\ref{2eqn2}) by demanding $\bar{\phi}_{1}=i\phi_{1}$ and $\bar{\phi}_{2}=i\phi_{2}$. For a real Clifford algebra $Cl(p,0)=Cl(0,q)$, our complex vector space becomes $V\bigotimes \mathbb{C}\simeq a\bigoplus\bar{a}$ 
\begin{equation}\label{2eqn3}
a_{i}=\frac{1}{2}(e_{i}+i\phi_{i})
\end{equation}
\begin{equation}\label{2eqn4}
a^{\dagger}_{i}=\frac{1}{2}(e_{i}-i\phi_{i})
\end{equation}
satisfying Eq.\ref{2eqn2}. Extending the metric linearly into the complex field gives the canonical commutation relations for fermionic oscillators 
\begin{align}\label{2eqn5}
    \{a_{i},a_{j}\}=0&&\{a^{\dagger}_{i},a^{\dagger}_{j}\}=0&&\{a_{i},a^{\dagger}_{j}\}=\delta_{ij}
\end{align}
as our maximally isotropic sub-spaces\cite{3cit1,cit16}. The basis space for $\mathbb{C}l(n)$ is then generated from combinations of $n/2$ basis elements $a_{i},a^{\dagger}_{i}$ satisfying Eq.(\ref{2eqn5}). Irreducible representations are given by minimal left ideals 
\begin{equation}\label{minide}
a^{\dagger}_{m_{1}}...a^{\dagger}_{m_{n}}\ket{\Omega}
\end{equation}
with $\ket{\Omega}$ behaving as an idempotent vacuum. For a review on idempotent operators and minimal left ideals in a Witt basis and real Clifford algebra formulation, please refer to the works\cite{witt1,3cit1,1cit16}. Our current approach is unique in that we can use these general isotropic sub-spaces Eq.(\ref{2eqn5}) to construct unitary symmetries without the automorphism of non-associative algebras\cite{cit14}.This approach allows us to explore the bivector algebra independently, without relying on the automorphism of non-associative algebras. One can show that our isotropic subspace exhibits a unitary Lie algebra structure by calculating the Lie bracket of bilinear products of the form $K_{ij}=a_{i}a^{\dagger}_{k}$ giving a Lie bracket
\begin{align}\label{2eqn6}
    [K_{ij},K_{km}]=[a_{i}a^{\dagger}_{j},a_{k}a^{\dagger}_{m}]&=a_{i}a^{\dagger}_{m}\delta_{jk}-a_{k}a^{\dagger}_{j}\delta_{im}
\end{align}
with $K_{ij}$ forming an embedding of $U(N)$ structure into our Clifford algebra. For a general bilinear operator $\bold{A}=a_{i}A_{ij}a^{\dagger}_{j}$  we can calculate its Lie bracket as 
\begin{align}\label{2eqn7}
    [\bold{A},\bold{B}]&=a_{i}[\bold{C}]_{ij}a^{\dagger}_{j}
\end{align}
allowing us to form a $U(N)$ Lie algebra. In the section to come, we will recover information on the real Clifford algebra lost during complexification by first imposing a unitary symmetry in the complex algebra Eq.(\ref{2eqn7}) and then studying its dual real algebra acquired by satisfying Eq.(\ref{2eqn22}) through the complex structure giving Eq.(\ref{2eqn3},\ref{2eqn4}). Notably, we observe a departure from the conventional approach seen in gauge theories, which introduce gauge groups as principal bundles over a fixed space-time\cite{3cit3}. This section introduces a distinctive perspective, the emergence of a specific vector space is constrained by the canonical commutation relation. Moreover, this vector space can be effectively correlated with the gauge group. This observation is intriguing because the formalism resembles aspects found in Kaluza-Klein theories. The introduction of extra dimensions, regulated by compactification methods, leads to the emergence of particle phenomena\cite{3cit2}\cite{1cit12}. 
\section{Unitary Symmetries} \label{sec:unitary}
\subsection{Weak Interactions}\label{sec:unitary1}
The first gauge group that we are interested in relates to the weak interactions. Using  Eq.(\ref{2eqn7}) we can immediately write down the generators of $SU(2)$ 
\begin{align}
    T^{1}=a_{1}a^{\dagger}_{2}-a^{\dagger}_{1}a_{2}&&T^{2}=-ia_{1}a^{\dagger}_{2}+ia_{2}a^{\dagger}_{1}&& T^{3}=a_{1}a^{\dagger}_{1}-\alpha_{2}\alpha^{\dagger}_{2}
\end{align}
The exponentiation of quadratic bivector elements
maps into $Spin(n)$\cite{Cliffbook2,1cit18}, forming a class of exceptional isomorphisms for $n\le 8$. To maintain consistency upon exponentiating operators, the generators that we are interested in are infinitesimal operator generators given of the form 
\begin{align}\label{4eqn111}
    T^{1}=ia_{1}a^{\dagger}_{2}-ia^{\dagger}_{1}a_{2}&&T^{2}=a_{1}a^{\dagger}_{2}-a_{2}a^{\dagger}_{1}&& T^{3}=ia_{1}a^{\dagger}_{1}-ia_{2}a^{\dagger}_{2}
\end{align}
Instead of using these generators to express our automorphism group in term of operators\cite{cit14,cit15,cit16}, we instead find their corresponding bivector generators. Our first indefinite signature of interest starts with $Cl(1,3)$ often denoted as the Space-Time algebra\cite{cit13,cit14,Cliffbook2} generated by the basis orthonormal basis vectors $\{\gamma_{0},\gamma_{1},\gamma_{2},\gamma_{3}\}$ given observation that $SU(2)$ demands a 16 dimensional space as explicitly seen in Eq.(\ref{2eqn3},\ref{2eqn4}). The ladder operators are given in a straightforward manner denoted as 
\begin{align}\label{4ladder1}
    a_{1}=\frac{1}{2}(\gamma_{0}+\gamma_{1})&&a_{1}^{\dagger}=\frac{1}{2}(\gamma_{0}-\gamma_{1})
 \end{align}
 \begin{align}\label{4ladder2} 
 a_{2}=\frac{1}{2}(i\gamma_{3}+\gamma_{2})&&a_{2}^{\dagger}=\frac{1}{2}(i\gamma_{3}-\gamma_{2})
\end{align}
where complexification occurs for one spatial component in order to satisfy Eq.(\ref{2eqn22}). At first glance, the interchange between spatial elements $\gamma_{i}\rightarrow\gamma_{j}$ leaves our anti-commutator invariant and it is rather unclear what choice from 3 possible exchanges is the correct one. It will turn out that the exchange does not matter when it comes to the Lie algbera given that we're leaving equations Eq.(\ref{2eqn5}) invariant. The corresponding bivector generators of Eq.(\ref{4eqn111}) are found to be 
\begin{align}\label{2lorentz}
T_{i}=\frac{1}{2}(\epsilon_{ijk}\gamma_{j}\gamma_{k}+i\gamma_{0}\gamma_{i})
\end{align} 
To analyze what these bivector generators represent, recall the Lie algebra for the Lorentz group is $\mathfrak{so}(1,3)\simeq sl(2,\mathbb{C})\oplus sl(2,\mathbb{C})$\cite{4cit1}, where the unitarian trick gives a one to one correspondence between special linear representations and special unitary representations of $\mathfrak{so}(1,3)_{\mathbb{C}}\simeq\mathfrak{su}(2)_{\mathbb{C}}\oplus\mathfrak{su}(2)_{\mathbb{C}}$. Our operator symmetry brought us in correspondence to $\mathfrak{su}(2)_{c}$ upon translating to its bivector form as a consequence of complexifying the Lorentz generators $\bold{K}$ and $\bold{M}$. The Lorentz generators are then easily identified as
\begin{align}\label{4eqn1}
\mathbf{K}_{1}=M_{01}=\gamma_{0}\gamma_{1}&&\mathbf{K}_{2}=M_{02}=\gamma_{0}\gamma_{2}&&\mathbf{K}_{3}=M_{03}=\gamma_{0}\gamma_{3}
\end{align}
\begin{align}\label{4eqn2}
\mathbf{M}_{1}=M_{23}=\gamma_{2}\gamma_{3}&&\mathbf{M}_{2}=M_{31}=\gamma_{3}\gamma_{1}&&\mathbf{M}_{3}=M_{12}=\gamma_{1}\gamma_{2}
\end{align}
with $\bold{K}_{i}$ forming non-unitary representations of $\mathfrak{so}(1,3)$ giving a non-compact representation upon exponentiating as a consequence of $\bold{K}_{i}^{2}=1$. On the other hand, $\bold{M}_{i}$ forms a unitary rotation subalgebra $\mathfrak{so}(3)$ of $\mathfrak{so}(1,3)$ where our bivector commutator bracket 
\begin{equation}
    [M_{\mu\nu}\times M_{\rho\sigma}]=\eta_{\nu\rho}M_{\mu\sigma}-\eta_{\mu\rho}M_{\nu\sigma}-\eta_{\sigma\mu}M_{\rho\nu}+\eta_{\sigma\nu}M_{\rho\mu}
\end{equation}
resembles the usual Lorentz commutation relation Eq.(\ref{2eqn}). Lorentz boost and generators can be identified with $\mathbb{R}^{3}$ by allowing the isomorphism $e_{i}\rightarrow\gamma_{0}\gamma_{i}$. From this isomorphism we can rewrite Eq.(\ref{4eqn1}) and Eq.(\ref{4eqn2}) to express boost as vectors $\mathbb{R}^{3}$ basis vectors and rotations as oriented planes of $\mathbb{R}^{3}$ independent of time. While the study of $Spin$ groups and their application to Lie algebras has been studied, deriving unitary symmetries in indefinite signatures through a Witt decomposition such as we offers an alternative. Usually one would derive special unitary groups in its bivector algebra by considering a doubling bivector $J=e_{j}\wedge f^{j}$ generating a $U(1)$ subgroup of $\mathfrak{su}(p,q)$ where $f^{j}$ is a anti-commuting second frame to $e_{j}$. Where one can then factor out $J$ to form a basis for $\mathfrak{su}(p,q)$\cite{witt2,witt1}. While this method is straightforward for Euclidean signatures it can be rather cumbersome for non-Euclidean signatures. The question on whether $Cl(1,3)$ is the underlying real algebra is answered by restricting the field of our Lie algebra to $\mathbb{R}$. This ties us to the generators $\bold{M}$ and $\bold{K}$ giving an Spin adjoint action of
\begin{equation}
    Spin(1,3)\simeq SL(2,\mathbb{C})
\end{equation}
upon exponentiation which is not the gauge symmetry that we are seeking. Although we arrived at a complex $\mathfrak{su}(2)_{\mathbb{C}}$ bivector representation, this will not always be the case. In fact the real algebra we're seeking will be the one where there is no need to restrict our field over $\mathbb{R}$.
\par
We now study the Clifford algebra $Cl(4,0)$ carrying a pure Euclidean signature. One can express the operator generators in terms of bivector elements by first defining
\begin{align}
    a_{1}=\frac{1}{2}(e_{1}+i\phi_{1})&&a^{\dagger}_{1}=\frac{1}{2}(e_{1}-i\phi_{1})
\end{align}
 \begin{align}
 a_{2}=\frac{1}{2}(e_{2}+i\phi_{2})&&a^{\dagger}_{2}=\frac{1}{2}(e_{2}-i\phi_{2})
\end{align}
In this case two spatial components are complexified satisfying Eq.(\ref{2eqn22})
\begin{align}\label{4eqnsu}
    T^{1}=\frac{1}{2}(e_{1}\phi_{2}-\phi_{1}e_{2})&&T^{2}=\frac{1}{2}(e_{1}e_{2}+\phi_{1}\phi_{2})
\end{align}
\begin{equation}\label{4eqnsu1}
 T^{3}=\frac{1}{2}(e_{1}\phi_{1}-e_{2}\phi_{2})
 \end{equation}
For $Cl(1,3)$, restricting Eq.(\ref{2lorentz}) over $\mathbb{R}$ brought us in correspondence with generators of $Spin(1,3)$(\ref{4eqn1}\ref{4eqn2}). This tells us ultimately that real bivector structures will fall under elements of the $Spin$ group, and its this realization that the $Spin$ group highly restricts the Clifford algebras and their basis spaceEq(\ref{3eqn}). Studying whether our bivector algebra is chiral or not can be answered through the manifestation of $\mathfrak{su}(2)_{L}$ given by the exceptional isomorphism
\begin{equation}\label{4eqn33}
    Spin(4)=SU(2)\times SU(2)
\end{equation}
As emphasized in the introduction, the bivector structure of a Geometric algebra allows us to write down the Lie algebra in a forward manner. This is mainly due to the fact that we work in a representation-free form using basis elements, which is exhibited in the following. We are interested in $\mathfrak{so}(4)$ adhering a commutator bracket 
\begin{equation}\label{4eqn4}
    [X_{ij}\times X_{km}]=\delta_{im}X_{jk}-\delta_{ik}X_{mj}-\delta_{jm}X_{ik}+\delta_{jk}X_{im}
\end{equation}
The six generators which satisfy this bracket are presented along with their commutator brackets 
\begin{align}
\mathcal{A}^{3}=e_{1}e_{2}&&\mathcal{A}^{1}=e_{2}e_{3}&&\mathcal{A}^{2}=e_{1}e_{3}
\end{align}
\begin{align}
\mathcal{B}^{1}=e_{1}e_{4}&&\mathcal{B}^{2}=e_{4}e_{2}&&\mathcal{B}^{3}=e_{3}e_{4}
\end{align}
\begin{align}
    [\mathcal{A}_{i}\times\mathcal{A}_{j}]=\epsilon_{ijk}\mathcal{A}_{k}&&[\mathcal{B}_{i}\times\mathcal{B}_{j}]=\epsilon_{ijk}\mathcal{A}_{k}&&[\mathcal{A}_{i}\times\mathcal{B}_{j}]=\epsilon_{ijk}\mathcal{B}_{k}
\end{align}
where both $\mathcal{A}_{i}$ and $\mathcal{B}_{i}$ are necessary to close the algebra. At the Lie algebra level, our exceptional isomorphism is expressed in the form $\mathfrak{so}(4)=\mathfrak{su}(2)\oplus\mathfrak{su}(2)$ where our two copies of $\mathfrak{su}(2)$ are given as
\begin{align}\label{construction}
   J^{\pm}_{i}=(A_{i}\pm B_{i})
\end{align}
satisfying $[J^{+},J^{-}]=0$. This form is suggestive in that we can define $\mathfrak{su}(2)_{R}$ and $\mathfrak{su}(2)_{L}$ Lie algebras by considering chiral projection operators $P_{L}/P_{R}$ acting on $\mathfrak{su}(2)$ elements from the exceptional isomorphism $Spin(3)\simeq SU(2)$, that is second grade elements of $Cl(3,0)$ Eq.(\ref{5eqn}) deducing that a projection operator of the form
\begin{equation}\label{4eqn3}
    P_{R/L}=\frac{1}{2}(1\pm e_{1}e_{2}\phi_{1}\phi_{2})
\end{equation}
We have thus arrived at left and right handed algebras by acting with a 4 dimensional projection operator on $Cl(3,0)$, where chirality emerges from the addition of a spatial dimension to our Euclidean signature. But we have also arrived at $SU(2)_{L}$ by starting off with a complex algebra and finding its real dual .In the study of $Cl(1,3)$ as the underlying algebra, we observed that our operator algebra brought us into $\mathfrak{su}(2)_{\mathbb{C}}$ Eq.(\ref{2lorentz}). On the other hand by considering $Cl(4,0)$ we were brought upon $\mathfrak{su}(2)_{L}$ Eq.(\ref{4eqn1},\ref{4eqn2}) already over the field $\mathbb{R}$, which is the gauge symmetry that we are looking. These remarks show that not every operator $\mathfrak{su}(2)$ algebra will map into a real $\mathfrak{su}(2)$ algebra, resembling the well known mathematical fact that not every Lie algebra gives the same Lie group. Even more surprising is the appearance of $\mathfrak{su}(2)_{L}$ as preference over $\mathfrak{su}(2)_{R}$ which is responsible for the partaking of left handed and right handed anti particles in the weak interactions. The isomorphism of $e_{i}\rightarrow\gamma_{0}\gamma_{i}$ extends our chiral projection operator Eq.(\ref{4eqn3}) onto $Cl(1,4)$ where we effectively included an extra spatial component to the Space-Time algebra. Similarly, the isomorphisms between $Cl(4,1)$ and $\mathbb{C}l(4)$ led to the work of \cite{1cit19,1cit20,1cit21} where basis ladder states are placed in correspondence with elements of $Cl(4,1)$ which is algebraic different than our implied $Cl(1,4)$. We notice that an odd number of basis vectors implies that not every Clifford algebra has an straight forward dual description of ladder operators given that our formulation relies on the even partitioning in the definition of null vectors Eq.(\ref{2eqn2}). This opens up the question whether spaces with odd elements can in general accommodate a ladder description by implementing compactifying methods via the complex structure as done in \cite{1cit12}to reduce the number of odd elements to even elements and explore the emergent physics from these inclusion of extra basis elements.
\par
Having discovered $Cl(4,0)$ as the appropriate underlying algebra prior to complexification, irreducible representations of the complex Clifford algebra follow directly from Eq.(\ref{minide}) with an overview referenced in section\ref{sec:complex}. A physical interpretation is that bosonic irreducible representations are infinite as a consequence of no nilpotency, where the nilpotency of our fermion ladder operators limits us at a finite number of states. It is then a forward manner to identify particle states Table~(\ref{tab:table1})
\begin{table}[H]
\caption{\label{tab:table1}
}
\begin{ruledtabular}
\begin{tabular}{cccc}
\textrm{Particle}&
\textrm{State}&
\multicolumn{1}{c}{\textrm{Irreducible Rep $SU(2)$}}&
\textrm{$U(1)_{Y}$ Charge}\\
\colrule
$\nu_{R}$ & $\ket{\Omega}$ & $\bold{1}$ & 0\\
$\nu_{L}$ & $a^{\dagger}_{1}\ket{\Omega}$ & $\bold{2}$ & $-1$\\
$e^{-}_{L}$ & $a^{\dagger}_{2}\ket{\Omega}$ & $\bold{2}$ & $-1$\\
$e^{-}_{R}$ &$a^{\dagger}_{1}a^{\dagger}_{2}\ket{\Omega}$  & $\bold{1}$ & $-2$\\
\end{tabular}
\end{ruledtabular}
\end{table}
\begin{table}[H]
\caption{\label{tab:table2}
}
\begin{ruledtabular}
\begin{tabular}{cccc}
\textrm{Particle}&
\textrm{State}&
\multicolumn{1}{c}{\textrm{Irreducible Rep $SU(2)$}}&
\textrm{$U(1)_{Y}$ Charge}\\
\colrule
$\bar{\nu}_{L}$ & $\ket{\Omega}^{*}$ & $\bold{\bar{1}}$ & $1$\\
$\bar{\nu}_{R}$ & $a_{1}\ket{\Omega}^{*}$ & $\bold{\bar{2}}$ & $0$\\
$e^{+}_{R}$ & $a_{2}\ket{\Omega}^{*}$ & $\bold{\bar{2}}$ & $2$\\
$e^{+}_{L}$ &$a_{1}a_{2}\ket{\Omega}^{*}$  & $\bold{\bar{1}}$ &$1$\\
\end{tabular}
\end{ruledtabular}
\end{table}
along with its conjugate Table~(\ref{tab:table2}) transforming under the unitary symmetries Eq.(\ref{4eqn111}) obtained from our general Witt decomposition. Our Clifford vacuum takes the form $\ket{\Omega}=a_{1}a_{2}a^{\dagger}_{2}a^{\dagger}_{1}$ and its conjugate $\ket{\Omega}^{*}=a^{\dagger}_{2}a^{\dagger}_{1}a_{1}a_{2}$ coming in agreement with the particle states described as minimal left ideals for the weak interactions as described in\cite{cit15,cit16}. The weak hyperchage is computed as the $U(1)_{Y}$ conserved quantity which can be easily deduces as the number operator $Y=-\frac{N}{2}$\cite{cit15} which commutes with all of the generators in Eq.(\ref{4eqn111}). We have arrived at the observation that the $\mathbb{C}\otimes Cl(4,0)$ provides the correct operator gauge symmetries in addition to preserving the bivector gauge symmetry upon translation. An additional indefinite signature of possible interest is $Cl(2,2)$ carrying an exceptional isomorphism $Spin(2,2)=SL(2,\mathbb{R})\times SL(2,\mathbb{R})$ recognized as the restricted conformal group, the non-compact version of $SO(4)=SU(2)\times SU(2)$ of $Cl(4,0)$. The restricted conformal lie algebra has the isomorphism $conf(1,1)\simeq \mathfrak{so}(2,2)$ generated by basis elements $\{M_{\mu\nu},K_{\mu},P_{\mu},D\}$ described by  Lorentz transformations, special conformal transformations,translations and dilations. Writing down the ladder operators for this signature satisfying Eq.(\ref{2eqn22}) will lead to a complex bivector algebra, given that the complex structure will carry over into the bivector representation. Restricting over $\mathbb{R}$ will give us representations of ${\mathfrak{so}(2,2)}$ eliminating $Cl(2,2)$ as a group for the weak interactions. As mentioned in section(\ref{sec:complex}) our procedure allowed us to arrive at particle representations and gauge symmetry independent of the the group automorphisms. We have shown that the ambiguity in not knowing what space to complexify can be eliminated by looking at the $Spin$ group when going from the operator algebra to the bivector algebra. Our general Witt decomposition in addition allowed us to construct unitary symmetries directly in their bivector form such as Eq.(\ref{2lorentz},\ref{4eqnsu},\ref{4eqnsu1}) without us having to construct the symmetry group explicitly as done in Eq.(\ref{construction}).
\subsection{Strong Interactions}
Following the outline of section(\ref{sec:unitary}) we would like to understand the space which admits an $SU(3)_{C}$ symmetry describing strong interactions. We start from our operator Lie algebra Eq(\ref{2eqn7}) admitting generators of the form  
\begin{align}\label{su31}
    T^{1}=F_{12}=ia_{1}a^{\dagger}_{2}-ia^{\dagger}_{1}a_{2}&&T^{2}=E_{12}=a_{1}a^{\dagger}_{2}+a^{\dagger}_{1}a_{2}
\end{align}
\begin{align}\label{su32}
    T^{3}=H_{1}=ia_{1}a^{\dagger}_{1}-ia_{2}a^{\dagger}_{2}&& T^{4}=F_{13}=ia_{1}a^{\dagger}_{3}-ia^{\dagger}_{1}a_{3}
\end{align}
\begin{align}\label{su33}
T^{5}=E_{13}=a_{1}a^{\dagger}_{3}+a^{\dagger}_{3}a_{1}&&T^{6}=F_{23}=ia_{2}a^{\dagger}_{3}-ia^{\dagger}_{3}a_{2}
\end{align}
\begin{align}\label{su34}
T^{7}=E_{23}=a_{2}a^{\dagger}_{3}+a^{\dagger}_{3}a_{2}&&T^{8}=\frac{1}{2\sqrt{3}}(ia_{1}a^{\dagger}_{1}+ia_{2}a^{\dagger}_{2}-ia_{3}a^{\dagger}_{3})
\end{align}
where we have already taken into account their infinitesimal variation as mentioned in section(\ref{sec:unitary1}) . By examining the ladder operators, we identify the requirement for a 64-dimensional space, requiring three separate sets of operators each satisfying Eq.(\ref{2eqn5}). As discussed in the previous section, an examination of the $Spin$ groups provides valuable insights into determining the suitable real Clifford, considering that the group acts as a constraint on the space. Applying this learned reasoning, we consider $Cl(6,0)$ as our primary algebra to start with due to its exceptional isomorphism
\begin{equation}
Spin(6)\simeq SU(4)
\end{equation}
This consideration is crucial, given the disparity between weak and strong interactions gauge groups. This disparity is the existence of an explicit relation between $Spin(4)$ and $SU(2)$ through $Spin(4)=SU(2)\times SU(2)$, while no such relation exists for strong interactions. In the earlier section(\ref{sec:unitary1}), our exploration led us to focus on $Cl(4,0)$ to obtain $Spin(4)\simeq SO(4)=SU(2)\times SU(2)$, yielding a chiral Lie algebra expressed as $J^{\pm}_{i}=(A_{i}\pm B_{i})$. This insight is particularly valuable given the existence of a subgroup and two canonical subgroups of $\mathfrak{su}(2)$ can be found within $\mathfrak{su}(3)$. One can then establish a natural extension into $\mathfrak{su}(3)$ if one can express $\mathfrak{su}(2)$ in terms of $Spin(6)$ generators. We find $\mathfrak{so}(6)$ bivector generators satisfying Eq.(\ref{4eqn4}) which takes the form 
\begin{align}
    \Gamma^{1}=\frac{1}{2}e_{1}e_{2}&&\Gamma^{2}=\frac{1}{2}e_{1}e_{3}&&\Gamma^{3}=\frac{1}{2}e_{1}\phi_{1}&&\Gamma^{4}=e_{1}\phi_{2}
\end{align}
\begin{align}
    \Gamma^{5}=\frac{1}{2}e_{1}\phi_{3}&&\Gamma^{6}=\frac{1}{2}e_{2}e_{3}&&\Gamma^{7}=\frac{1}{2}e_{2}\phi_{1}&&\Gamma^{8}=e_{2}\phi_{2}
\end{align}
\begin{align}
    \Gamma^{9}=\frac{1}{2}e_{2}\phi_{3}&&\Gamma^{10}=\frac{1}{2}e_{3}\phi_{1}&&\Gamma^{11}=\frac{1}{2}e_{3}\phi_{2}&&\Gamma^{12}=\frac{1}{2}e_{3}\phi_{3}
\end{align}
\begin{align}
    \Gamma^{13}=\frac{1}{2}\phi_{1}\phi_{2}&&\Gamma^{14}=\frac{1}{2}\phi_{1}\phi_{3}&&\Gamma^{15}=\frac{1}{2}\phi_{2}\phi_{3}
\end{align}
It is then forward to see that $\mathfrak{su}(2)_{L}$ Eq.(\ref{4eqnsu},\ref{4eqnsu1}) can be a generated from $Spin(6)$ as  
\begin{align}
   T^{1}=\Gamma^{4}-\Gamma^{7}&&T^{2}=\Gamma^{1}+\Gamma^{13}&&T^{3}=\Gamma^{3}-\Gamma^{8}
\end{align}
It then follows that the remaining $\mathfrak{so}(6)$ generators can be extended to form the remaining 5 generators of $\mathfrak{su}(3)$ via
\begin{align}
   T^{4}=\Gamma^{5}-\Gamma^{10}&&T^{5}=\Gamma^{2}+\Gamma^{14}&&T^{6}=\Gamma^{9}+\Gamma^{11}
\end{align}
\begin{align}
   T^{7}=\Gamma^{6}+\Gamma^{15}&&T^{8}=\frac{1}{\sqrt{3}}(\Gamma^{3}+\Gamma^{8}-2\Gamma^{12})
\end{align}
coming into agreement of those found through by using the appropiate decomposition satisfying Eq.(\ref{2eqn3},\ref{2eqn4}) to find the their bivector dual. Given that we were able to find $\mathfrak{su}(3)$ within $Spin(6)$ as a bivector algebra over $\mathbb{R}$, we conclude that the extension of Euclidean space from $\mathbb{R}^{4}$ to $\mathbb{R}^{6}$ facilitates the inclusion of strong interaction, offering an alternative derivation to that in\cite{witt2}. The algebra at hand can be questioned given that color gauge symmetry does not exhibit chiral behavior, not allowing one to compute a projection operator such as Eq.(\ref{4eqn3}). From a represenation perspective there still lies the questions of having leptons in our particle spectrum when considering pure $SU(3)$ gauge theory. It is known that fermions transform under the fundamental representation of a gauge group in the study of quantum field theory, therefore the inclusion of leptons in $SU(3)$ bivector invariance but not quarks in $SU(2)$ bivector invariance is peculiar. From a group theoretical perspective, this observation might be comprehended as a result of the unbroken gauge symmetries $U(1)_{em}\times SU(3){c}$ emerging from $SU(4)$. We identify $SU(4)$ as the Pati-Salam group, highlighting the inherent tendency of unification groups to emerge seamlessly in the exploration of geometric algebras, facilitated by the isomorphisms of the $Spin$ group. Having identified the correct Clifford algebra at hand through Spin arguments, the nilpotency property of our fermionic ladder operators allow for a finite number of representations allowing for the identification of our particle states given by Table~\ref{tab:table3}
\begin{table}[h]
\caption{\label{tab:table3}
}
\begin{ruledtabular}
\begin{tabular}{cccc}
\textrm{Particle}&
\textrm{State}&
\multicolumn{1}{c}{\textrm{Irreducible Rep SU(3)}}&
\textrm{$U(1)_{em}$ Charge}\\
\colrule
$\nu$ & $\ket{\Omega}$ & $\bold{1}$ & 0\\
$\bar{d}_{i}$ & $a^{\dagger}_{i}\ket{\Omega}$ & $\bold{\bar{3}}$ & $1/3$\\
$u_{k}$ & $a^{\dagger}_{i}a^{\dagger}_{j}\ket{\Omega}$ & $\bold{3}$ & $2/3$\\
$e^{+}$ & $a^{\dagger}_{1}a^{\dagger}_{2}a^{\dagger}_{3}\ket{\Omega}$ & $\bold{1}$ & 1\\
\end{tabular}
\end{ruledtabular}
\end{table}
along with its conjugate Table~(\ref{tab:table4})
\begin{table}[h]
\caption{\label{tab:table4}
}
\begin{ruledtabular}
\begin{tabular}{cccc}
\textrm{Particle}&
\textrm{State}&
\multicolumn{1}{c}{\textrm{Irreducible Rep SU(3)}}&
\textrm{$U(1)_{em}$ Charge}\\
\colrule
$\bar{\nu}$ & $\ket{\Omega}^{*}$ & $\bold{1}$ & 0\\
$d_{i}$ & $a_{i}\ket{\Omega}^{*}$ & $\bold{3}$ & $-1/3$\\
$\bar{u}_{k}$ & $a_{i}a_{j}\ket{\Omega}^{*}$ & $\bold{\bar{3}}$ & $-2/3$\\
$e^{-}$ &$a_{1}a_{2}a_{3}\ket{\Omega}^{*}$  & $\bold{1}$ & -1\\
\end{tabular}
\end{ruledtabular}
\end{table}
transforming under the unitary symmetries Eq.(\ref{su31},\ref{su32},\ref{su33},\ref{su34}) obtained from our general Witt decomposition, coming in agreement with the particle states described as ideals of $(\mathbb{C}\otimes \mathbb{O})_{L}$ for the strong interactions\cite{cit14,cit15} with with $\ket{\Omega}=a_{1}a_{2}a_{3}a^{\dagger}_{3}a^{\dagger}_{2}a^{\dagger}_{1}$. The charge is computed as the $U(1)_{em}$ conserved quantity corresponding to $SU(3)$, which can be easily deduces as the number operator $Q=\frac{N}{3}$ commuting with all of the generators of the operator $SU(3)$\cite{cit14,cit15}.We have again show that we can represent particle states as complexified elements of a Euclidean space. Working in a complex space facilities the transformation as states but also provides the ability to express our elements in a compact form. We can express a spinor as $\Psi=\psi^{\alpha}\Omega_{\alpha}$ where $\Omega_{\alpha}$ represent the vacuum ideal and its conjugate. This can be useful in the study of Spin gauge fields as described in references\cite{1cit_3,1cit_4}. In our last sections we discovered that the correct gauge symmetry over $\mathbb{R}$ is unique for a given $D$ dimensional space we are interested in. Despite this statement, they're various signatures that one can quickly discard by applying the learned remarks throughout this article. One could've considered the indefinite signature $Cl(3,3)$ which carries and accidental isomorphism 
\begin{equation}
    Spin(3,3)\simeq SL(4,\mathbb{R})
\end{equation}
which is isomorphic to the conformal group $SO(3,3)$ carrying 15 generators compromised of the basis elements $\{M_{\mu\nu},K_{\mu},P_{\mu},D\}$ as described in section(\ref{sec:unitary1}). In the context of unitary symmetries it has been studied that $SO(3,3)$ contains two complexified subgroups of $\mathfrak{su}(2)_{\mathbb{C}}$ generated by $SO(1,3)$ and $SO(3,1)$\cite{4cit2}. Given again that one can usually come across $SU(2)$ as a subgroup and canonnical subgroup then any $SU(3)$ that is found will be complexified.  This is also verified by writing down the ladder operators for this signature satisfying Eq.(\ref{2eqn22}) where the complex structure will survive into the bivector representation from our $SU(3)$ operator generators, where restricting over $\mathbb{R}$ will give us representations of ${\mathfrak{so}(3,3)}$ eliminating $Cl(3,3)$. Similarly one can consider the indefinite signature $Cl(1,5)$ which contains the accidental isomorphism of the form 
\begin{equation}\label{last1}
Spin(1,5)\simeq SL(2,\mathbb{H})
\end{equation}
Mathematically we arrive at this algebra through the inclusion of two spatial dimensions to our Space-Time algebra from section(\ref{sec:unitary1}). Complexifying leads to a set of ladder operators with one set of operator remaining over the reals as in Eq.(\ref{4ladder1}). This again leads to complex nature of the underlying symmetry as we learned when studying $Cl(1,3)$. Lastly one can consider the indefinite signature where $Cl(2,4)$ exhibiting the relation 
\begin{equation}\label{last2}
Spin(2,4)\simeq SU(2,2)
\end{equation}
which one recognizes in the study of twistors in flat space-time belonging to the conformal double covering group $O(2,4)$ describing the momentum and angular momentum structure of zero-rest-mass particle, with a twistor approach in Clifford algebras studied in\cite{4cit4}. Finding the operator correspondence for this algebra would leave out two operator degrees of freedom uncomplexified which again restricting over $\mathbb{R}$ gives us the bivector generators of $SU(2,2)$, illustrating the fundamental observation throughout this paper in showing the relation between the underlying geometry of our space and the symmetries that its allowed to admit in or study of the Standard Model. Although these $Spin$ groups are not of our current interest, they play an important role for example in the study of conformal structures\cite{4cit3,4cit4}. In section\ref{sec:unitary1} it was mentioned that $Cl(1,4)$ and $Cl(4,1)$ have different algebraic structures despite their $Spin$ groups being equivalent, which also applies to\ref{last1}\ref{last2}. From this point of view, there would have to be a mechanism which would allows the distinction between these two algebras. This brings up another reason to consider the underlying real Clifford algebra to be of Euclidean signature where $Cl(p,0)$ and$ Cl(0,q)$ are equivalent and there would be no distinction to account to the Lie algebra. 
\section{Conclusion}\label{sec:conclusion}
Aims of this article were to establish a deeper understanding between complex Clifford algebra emerging from left adjoint actions as constructed in\cite{cit14,cit15,cit16,3cit1} and its underlying real Clifford algebra. We established the remark that one can use a generalized Witt decomposition\cite{witt1,witt2} where a complex structure is introduced to construct unitary symmetries independent of the automorphism groups of non-associative algebras to constraint our vector space. Establishment of this statement introduces a distinctive perspective, the emergence of a specific vector space and its correlation to gauge groups. We show that the ambiguity in not knowing what space to complexify can be eliminated by looking at the $Spin$ groups. We do this by analyzing various indefinite signatures to find the appropriate Lie algebra. We solidify this statement through the study of $Spin(4)$, discovering the extension of a Euclidean signature Clifford algebra from $D=3\rightarrow D=4$ generating the weak interaction gauge group. In constructing our chiral algebra in $Cl(4,0)$, we observed natures preference in $\mathfrak{su}(2)_{L}$ over $\mathfrak{su}(2)_{R}$ when going from the operator algebra to its bivector algebra. Particle states are then identified as irreducible representations of $\mathbb{C}\otimes Cl(4,0)$\cite{witt1}. We consider the extension from Euclidean signature Clifford algebra $D=3\rightarrow D=6$ allowing the inclusion of strong interactions by considering $Spin(6)$ as a means to acquire leptons and quarks charged under $\mathbb{C}\otimes Cl(6,0)$. The emergence of the Pati-Salam group resembles aspects found in Kaluza-Klein theories and unification models. The connection explored in this article between a real Clifford algebra and its Complex Clifford algebra through the Spin groups and general Witt decomposition will aid in understanding how an operator based Clifford action\cite{1cit_1,1cit_2,1cit_3,1cit_4} for example can be translated into its bivector field theory representation form\cite{1cit9,1cit11,1cit16}.This will allow the exploration of extra dimensional aspects of $Cl(6,0)$ and $Cl(4,0)$ as a bivector field theory. Understanding the implication that our $Cl(1,4)$ left and right projection operator has on $Cl(1,4)$ and whether one can find a dual ladder description by compactifying the additional dimension through its complex structure are worth studying. Our general Witt decomposition allows us to construct unitary symmetry directly without having to construct the symmetry group explicitly from first principles. This can open roads in being able to study a variety of Lie algebras in terms of their bivector structures. Results of this articles and future aims highlight the deep interplay between geometry and physics in fundamental interactions, where Clifford algebras offer a concise and versatile formalism from which this can be explored. 
\section*{Acknowledgements} \label{sec:acknowledgements}
The author would like to give special thanks to his advisor Thomas Klähn at CSULB without whom this work would not have been possible. He would also like to give Kathryn McCormick at the math department at CSULB thanks for helpful suggestions on this manuscript. This work has been supported by the Google Summer Research Assistantship 2022 and Google Summer Research Assistantship 2023.
%\printbibliography
\bibliography{example}

\begin{thebibliography}{49}%
\makeatletter
\providecommand \@ifxundefined [1]{%
 \@ifx{#1\undefined}
}%
\providecommand \@ifnum [1]{%
 \ifnum #1\expandafter \@firstoftwo
 \else \expandafter \@secondoftwo
 \fi
}%
\providecommand \@ifx [1]{%
 \ifx #1\expandafter \@firstoftwo
 \else \expandafter \@secondoftwo
 \fi
}%
\providecommand \natexlab [1]{#1}%
\providecommand \enquote  [1]{``#1''}%
\providecommand \bibnamefont  [1]{#1}%
\providecommand \bibfnamefont [1]{#1}%
\providecommand \citenamefont [1]{#1}%
\providecommand \href@noop [0]{\@secondoftwo}%
\providecommand \href [0]{\begingroup \@sanitize@url \@href}%
\providecommand \@href[1]{\@@startlink{#1}\@@href}%
\providecommand \@@href[1]{\endgroup#1\@@endlink}%
\providecommand \@sanitize@url [0]{\catcode `\\12\catcode `\$12\catcode `\&12\catcode `\#12\catcode `\^12\catcode `\_12\catcode `\%12\relax}%
\providecommand \@@startlink[1]{}%
\providecommand \@@endlink[0]{}%
\providecommand \url  [0]{\begingroup\@sanitize@url \@url }%
\providecommand \@url [1]{\endgroup\@href {#1}{\urlprefix }}%
\providecommand \urlprefix  [0]{URL }%
\providecommand \Eprint [0]{\href }%
\providecommand \doibase [0]{https://doi.org/}%
\providecommand \selectlanguage [0]{\@gobble}%
\providecommand \bibinfo  [0]{\@secondoftwo}%
\providecommand \bibfield  [0]{\@secondoftwo}%
\providecommand \translation [1]{[#1]}%
\providecommand \BibitemOpen [0]{}%
\providecommand \bibitemStop [0]{}%
\providecommand \bibitemNoStop [0]{.\EOS\space}%
\providecommand \EOS [0]{\spacefactor3000\relax}%
\providecommand \BibitemShut  [1]{\csname bibitem#1\endcsname}%
\let\auto@bib@innerbib\@empty
%</preamble>
\bibitem [{\citenamefont {Georgi}\ and\ \citenamefont {Glashow}(1974)}]{Georgi:1974sy}%
  \BibitemOpen
  \bibfield  {author} {\bibinfo {author} {\bibfnamefont {H.}~\bibnamefont {Georgi}}\ and\ \bibinfo {author} {\bibfnamefont {S.~L.}\ \bibnamefont {Glashow}},\ }\bibfield  {title} {\bibinfo {title} {{Unity of All Elementary Particle Forces}},\ }\href@noop {} {\bibfield  {journal} {\bibinfo  {journal} {Phys. Rev. Lett.}\ }\textbf {\bibinfo {volume} {32}} (\bibinfo {year} {1974})}\BibitemShut {NoStop}%
\bibitem [{\citenamefont {M.}\ and\ \citenamefont {Gursey}(1973)}]{cit6}%
  \BibitemOpen
  \bibfield  {author} {\bibinfo {author} {\bibfnamefont {G.}~\bibnamefont {M.}}\ and\ \bibinfo {author} {\bibfnamefont {F.}~\bibnamefont {Gursey}},\ }\bibfield  {title} {\bibinfo {title} {{Quark structure and the octonions}},\ }\href@noop {} {\bibfield  {journal} {\bibinfo  {journal} {J. Math. Phys}\ }\textbf {\bibinfo {volume} {14}} (\bibinfo {year} {1973})}\BibitemShut {NoStop}%
\bibitem [{\citenamefont {M.}\ and\ \citenamefont {Gursey}(1974)}]{cit7}%
  \BibitemOpen
  \bibfield  {author} {\bibinfo {author} {\bibfnamefont {G.}~\bibnamefont {M.}}\ and\ \bibinfo {author} {\bibfnamefont {F.}~\bibnamefont {Gursey}},\ }\bibfield  {title} {\bibinfo {title} {{Quark statistics and octonions}},\ }\href@noop {} {\bibfield  {journal} {\bibinfo  {journal} {Phys. Rev. D}\ }\textbf {\bibinfo {volume} {19}} (\bibinfo {year} {1974})}\BibitemShut {NoStop}%
\bibitem [{\citenamefont {Dixon}(2004)}]{cit4}%
  \BibitemOpen
  \bibfield  {author} {\bibinfo {author} {\bibfnamefont {G.}~\bibnamefont {Dixon}},\ }\bibfield  {title} {\bibinfo {title} {{Division algebras: family replication}},\ }\href@noop {} {\bibfield  {journal} {\bibinfo  {journal} {J. Math. Phys}\ }\textbf {\bibinfo {volume} {45}} (\bibinfo {year} {2004})}\BibitemShut {NoStop}%
\bibitem [{\citenamefont {Furey}(2015)}]{cit14}%
  \BibitemOpen
  \bibfield  {author} {\bibinfo {author} {\bibfnamefont {C.}~\bibnamefont {Furey}},\ }\bibfield  {title} {\bibinfo {title} {Charge quantization from a number operator},\ }\href@noop {} {\bibfield  {journal} {\bibinfo  {journal} {Phys. Lett. B}\ }\textbf {\bibinfo {volume} {742}} (\bibinfo {year} {2015})}\BibitemShut {NoStop}%
\bibitem [{\citenamefont {Dubois-Violette}(2016)}]{cittt3}%
  \BibitemOpen
  \bibfield  {author} {\bibinfo {author} {\bibfnamefont {M.}~\bibnamefont {Dubois-Violette}},\ }\bibfield  {title} {\bibinfo {title} {Exceptional quantum geometry and particle physics},\ }\href@noop {} {\bibfield  {journal} {\bibinfo  {journal} {Nucl. Phys. B}\ }\textbf {\bibinfo {volume} {912}} (\bibinfo {year} {2016})}\BibitemShut {NoStop}%
\bibitem [{\citenamefont {Dubois-Violette}\ and\ \citenamefont {Todorov}(2019)}]{cit2}%
  \BibitemOpen
  \bibfield  {author} {\bibinfo {author} {\bibfnamefont {M.}~\bibnamefont {Dubois-Violette}}\ and\ \bibinfo {author} {\bibfnamefont {I.}~\bibnamefont {Todorov}},\ }\bibfield  {title} {\bibinfo {title} {{Exceptional quantum geometry and particle physics II}},\ }\href@noop {} {\bibfield  {journal} {\bibinfo  {journal} {Nucl. Phys. B}\ }\textbf {\bibinfo {volume} {938}} (\bibinfo {year} {2019})}\BibitemShut {NoStop}%
\bibitem [{\citenamefont {Dubois-Violette}\ and\ \citenamefont {Todorov}(2018)}]{cit3}%
  \BibitemOpen
  \bibfield  {author} {\bibinfo {author} {\bibfnamefont {M.}~\bibnamefont {Dubois-Violette}}\ and\ \bibinfo {author} {\bibfnamefont {I.}~\bibnamefont {Todorov}},\ }\bibfield  {title} {\bibinfo {title} {{Deducing the symmetry of the standard model from the automorphism and structure groups of the exceptional Jordan algebra}},\ }\href@noop {} {\bibfield  {journal} {\bibinfo  {journal} {Int. J. Mod. Phys. A}\ }\textbf {\bibinfo {volume} {33}} (\bibinfo {year} {2018})}\BibitemShut {NoStop}%
\bibitem [{\citenamefont {Furey}(2018{\natexlab{a}})}]{cit888}%
  \BibitemOpen
  \bibfield  {author} {\bibinfo {author} {\bibfnamefont {C.}~\bibnamefont {Furey}},\ }\bibfield  {title} {\bibinfo {title} {{Three generations, two unbroken gauge symmtries, and one eight-dimensional algebra}},\ }\href@noop {} {\bibfield  {journal} {\bibinfo  {journal} {Phys. Lett. B}\ }\textbf {\bibinfo {volume} {785}} (\bibinfo {year} {2018}{\natexlab{a}})}\BibitemShut {NoStop}%
\bibitem [{\citenamefont {Gillard}\ and\ \citenamefont {Gresnigt}(2019)}]{cit12}%
  \BibitemOpen
  \bibfield  {author} {\bibinfo {author} {\bibfnamefont {A.~B.}\ \bibnamefont {Gillard}}\ and\ \bibinfo {author} {\bibfnamefont {N.~G.}\ \bibnamefont {Gresnigt}},\ }\bibfield  {title} {\bibinfo {title} {Three fermion generations with two unbroken gauge symmetries from the complex sedenions},\ }\href@noop {} {\bibfield  {journal} {\bibinfo  {journal} {Eur. Phys. J.C}\ }\textbf {\bibinfo {volume} {79}} (\bibinfo {year} {2019})}\BibitemShut {NoStop}%
\bibitem [{\citenamefont {Gresnigt}\ \emph {et~al.}(2023)\citenamefont {Gresnigt}, \citenamefont {Gourlay},\ and\ \citenamefont {Varma}}]{cit13}%
  \BibitemOpen
  \bibfield  {author} {\bibinfo {author} {\bibfnamefont {N.}~\bibnamefont {Gresnigt}}, \bibinfo {author} {\bibfnamefont {L.}~\bibnamefont {Gourlay}},\ and\ \bibinfo {author} {\bibfnamefont {A.}~\bibnamefont {Varma}},\ }\bibfield  {title} {\bibinfo {title} {Three generations of colored fermions with $s_{3}$ family symmetry from caley-dickson sedenions},\ }\href@noop {} {\bibfield  {journal} {\bibinfo  {journal} {Eur. Phys. J.C}\ }\textbf {\bibinfo {volume} {83}} (\bibinfo {year} {2023})}\BibitemShut {NoStop}%
\bibitem [{\citenamefont {Manogue}\ and\ \citenamefont {Dray}(2010)}]{cit1}%
  \BibitemOpen
  \bibfield  {author} {\bibinfo {author} {\bibfnamefont {C.~A.}\ \bibnamefont {Manogue}}\ and\ \bibinfo {author} {\bibfnamefont {T.}~\bibnamefont {Dray}},\ }\bibfield  {title} {\bibinfo {title} {{Octonions, E6, and Particle Physics}},\ }\href@noop {} {\bibfield  {journal} {\bibinfo  {journal} {J. Phys. Conf. Ser.}\ }\textbf {\bibinfo {volume} {254}} (\bibinfo {year} {2010})}\BibitemShut {NoStop}%
\bibitem [{\citenamefont {M}(1976)}]{cit8}%
  \BibitemOpen
  \bibfield  {author} {\bibinfo {author} {\bibfnamefont {G.}~\bibnamefont {M}},\ }\bibfield  {title} {\bibinfo {title} {{Octonionic Hilbert spaces, the Poincare group and SU(3)}},\ }\href@noop {} {\bibfield  {journal} {\bibinfo  {journal} {J. Math. Phys.}\ }\textbf {\bibinfo {volume} {17}} (\bibinfo {year} {1976})}\BibitemShut {NoStop}%
\bibitem [{\citenamefont {Khalfan}\ and\ \citenamefont {Kurt}(2017)}]{cit10}%
  \BibitemOpen
  \bibfield  {author} {\bibinfo {author} {\bibfnamefont {C.~B. S. C. Y. G.~A.}\ \bibnamefont {Khalfan}}\ and\ \bibinfo {author} {\bibfnamefont {L.}~\bibnamefont {Kurt}},\ }\bibfield  {title} {\bibinfo {title} {Revisiting the role of octonions in hadronic physics},\ }\href@noop {} {\bibfield  {journal} {\bibinfo  {journal} {Phys. Part. Nuclei LetterT}\ }\textbf {\bibinfo {volume} {14}} (\bibinfo {year} {2017})}\BibitemShut {NoStop}%
\bibitem [{\citenamefont {Gresnigt}(2018)}]{cit11}%
  \BibitemOpen
  \bibfield  {author} {\bibinfo {author} {\bibfnamefont {N.~G.}\ \bibnamefont {Gresnigt}},\ }\bibfield  {title} {\bibinfo {title} {Braids, normed division algebras, and standard model symmetries},\ }\href@noop {} {\bibfield  {journal} {\bibinfo  {journal} {Phys. Lett. B}\ }\textbf {\bibinfo {volume} {783}} (\bibinfo {year} {2018})}\BibitemShut {NoStop}%
\bibitem [{\citenamefont {Gresnigt}(2020)}]{citnew1}%
  \BibitemOpen
  \bibfield  {author} {\bibinfo {author} {\bibfnamefont {N.~G.}\ \bibnamefont {Gresnigt}},\ }\bibfield  {title} {\bibinfo {title} {A topological model of composite preons from the minimal ideals of two clifford algebras},\ }\href@noop {} {\bibfield  {journal} {\bibinfo  {journal} {Phys. Lett. B}\ }\textbf {\bibinfo {volume} {808}} (\bibinfo {year} {2020})}\BibitemShut {NoStop}%
\bibitem [{\citenamefont {Furey}(2018{\natexlab{b}})}]{cit15}%
  \BibitemOpen
  \bibfield  {author} {\bibinfo {author} {\bibfnamefont {C.}~\bibnamefont {Furey}},\ }\bibfield  {title} {\bibinfo {title} {$su(3)_{C}\times su(2)_{L}\times u(1)_{Y}(\times u(1)_{X})$ as a symmetry of division algebraic ladder operators},\ }\href@noop {} {\bibfield  {journal} {\bibinfo  {journal} {Eur. Phys. J.C}\ }\textbf {\bibinfo {volume} {78}} (\bibinfo {year} {2018}{\natexlab{b}})}\BibitemShut {NoStop}%
\bibitem [{\citenamefont {Furey}(2016)}]{cit16}%
  \BibitemOpen
  \bibfield  {author} {\bibinfo {author} {\bibfnamefont {C.}~\bibnamefont {Furey}},\ }\emph {\bibinfo {title} {Standard Model physics from abstract mathematical objects ?}},\ \href@noop {} {Ph.D. thesis},\ \bibinfo  {school} {Univerisity of Waterloo} (\bibinfo {year} {2016})\BibitemShut {NoStop}%
\bibitem [{\citenamefont {F.~Brackx}\ and\ \citenamefont {Soucek}(2011)}]{witt1}%
  \BibitemOpen
  \bibfield  {author} {\bibinfo {author} {\bibfnamefont {H.~D.~S.}\ \bibnamefont {F.~Brackx}}\ and\ \bibinfo {author} {\bibfnamefont {V.}~\bibnamefont {Soucek}},\ }\bibfield  {title} {\bibinfo {title} {On the structure of complex clifford algebras},\ }\href@noop {} {\bibfield  {journal} {\bibinfo  {journal} {Adv. Appl. Clifford algebras}\ }\textbf {\bibinfo {volume} {21}} (\bibinfo {year} {2011})}\BibitemShut {NoStop}%
\bibitem [{\citenamefont {C.~Doran}\ and\ \citenamefont {Acker}(1993)}]{witt2}%
  \BibitemOpen
  \bibfield  {author} {\bibinfo {author} {\bibfnamefont {F.~S.}\ \bibnamefont {C.~Doran}, \bibfnamefont {D.~Hestenes}}\ and\ \bibinfo {author} {\bibfnamefont {N.~V.}\ \bibnamefont {Acker}},\ }\bibfield  {title} {\bibinfo {title} {Lie groups as spin groups},\ }\href@noop {} {\bibfield  {journal} {\bibinfo  {journal} {J. Math. Phys}\ }\textbf {\bibinfo {volume} {34}} (\bibinfo {year} {1993})}\BibitemShut {NoStop}%
\bibitem [{\citenamefont {Garling}(2011)}]{Cliffbook1}%
  \BibitemOpen
  \bibfield  {author} {\bibinfo {author} {\bibfnamefont {D.}~\bibnamefont {Garling}},\ }\href@noop {} {\emph {\bibinfo {title} {Clifford Algebras: An Introduction}}}\ (\bibinfo  {publisher} {Cambridge University Press},\ \bibinfo {year} {2011})\BibitemShut {NoStop}%
\bibitem [{\citenamefont {Lounesto}(1997)}]{Cliffbook3}%
  \BibitemOpen
  \bibfield  {author} {\bibinfo {author} {\bibfnamefont {P.}~\bibnamefont {Lounesto}},\ }\href@noop {} {\emph {\bibinfo {title} {Clifford Algebras and Spinors}}}\ (\bibinfo  {publisher} {Cambridge University Press},\ \bibinfo {year} {1997})\BibitemShut {NoStop}%
\bibitem [{\citenamefont {C.~Doran}(2003)}]{Cliffbook2}%
  \BibitemOpen
  \bibfield  {author} {\bibinfo {author} {\bibfnamefont {A.~L.}\ \bibnamefont {C.~Doran}},\ }\href@noop {} {\emph {\bibinfo {title} {Geometric Algebra for Physicists}}}\ (\bibinfo  {publisher} {Cambridge University Press},\ \bibinfo {year} {2003})\BibitemShut {NoStop}%
\bibitem [{\citenamefont {Hestenes}\ and\ \citenamefont {Gurtler.}(1975)}]{1cit10}%
  \BibitemOpen
  \bibfield  {author} {\bibinfo {author} {\bibfnamefont {D.}~\bibnamefont {Hestenes}}\ and\ \bibinfo {author} {\bibfnamefont {R.}~\bibnamefont {Gurtler.}},\ }\bibfield  {title} {\bibinfo {title} {Consistency in the formulation of the dirac, pauli and schr{\"o}dinger theories},\ }\href@noop {} {\bibfield  {journal} {\bibinfo  {journal} {J. Math. Phys.}\ }\textbf {\bibinfo {volume} {16}} (\bibinfo {year} {1975})}\BibitemShut {NoStop}%
\bibitem [{\citenamefont {Trayling}(1993)}]{1cit12}%
  \BibitemOpen
  \bibfield  {author} {\bibinfo {author} {\bibfnamefont {G.}~\bibnamefont {Trayling}},\ }\bibfield  {title} {\bibinfo {title} {Geometric approach to the standard model},\ }\href@noop {} {\bibfield  {journal} {\bibinfo  {journal} {arXiv:hep-th/9912231}\ } (\bibinfo {year} {1993})}\BibitemShut {NoStop}%
\bibitem [{\citenamefont {Lasenby}\ and\ \citenamefont {Gull}(1993)}]{1cit13}%
  \BibitemOpen
  \bibfield  {author} {\bibinfo {author} {\bibfnamefont {C.~D.~A.}\ \bibnamefont {Lasenby}}\ and\ \bibinfo {author} {\bibfnamefont {S.}~\bibnamefont {Gull}},\ }\bibfield  {title} {\bibinfo {title} {States and operators in the spacetime algebra},\ }\href@noop {} {\bibfield  {journal} {\bibinfo  {journal} {Found. Phys.}\ }\textbf {\bibinfo {volume} {23}} (\bibinfo {year} {1993})}\BibitemShut {NoStop}%
\bibitem [{\citenamefont {Hestenes}(1966)}]{1cit14}%
  \BibitemOpen
  \bibfield  {author} {\bibinfo {author} {\bibfnamefont {D.}~\bibnamefont {Hestenes}},\ }\href@noop {} {\emph {\bibinfo {title} {Space--Time Algebra}}}\ (\bibinfo  {publisher} {Gordon and Breach},\ \bibinfo {year} {1966})\BibitemShut {NoStop}%
\bibitem [{\citenamefont {Lounesto}(2001)}]{1cit15}%
  \BibitemOpen
  \bibfield  {author} {\bibinfo {author} {\bibfnamefont {P.}~\bibnamefont {Lounesto}},\ }\href@noop {} {\emph {\bibinfo {title} {Clifford Algebras and Spinors}}}\ (\bibinfo  {publisher} {Cambridge University Press},\ \bibinfo {year} {2001})\BibitemShut {NoStop}%
\bibitem [{\citenamefont {Lu}(2011)}]{1cit17}%
  \BibitemOpen
  \bibfield  {author} {\bibinfo {author} {\bibfnamefont {W.}~\bibnamefont {Lu}},\ }\bibfield  {title} {\bibinfo {title} {Yang-mills interactions and gravity in terms of clifford algebra},\ }\href@noop {} {\bibfield  {journal} {\bibinfo  {journal} {Adv. Appl. Clifford algebras}\ }\textbf {\bibinfo {volume} {21}} (\bibinfo {year} {2011})}\BibitemShut {NoStop}%
\bibitem [{\citenamefont {McClellan}(2021)}]{1cit19}%
  \BibitemOpen
  \bibfield  {author} {\bibinfo {author} {\bibfnamefont {G.~E.}\ \bibnamefont {McClellan}},\ }\bibfield  {title} {\bibinfo {title} {Operators and field equations in the electroweak sector of particle physics},\ }\href@noop {} {\bibfield  {journal} {\bibinfo  {journal} {Adv. Appl. Clifford algebras}\ }\textbf {\bibinfo {volume} {31}} (\bibinfo {year} {2021})}\BibitemShut {NoStop}%
\bibitem [{\citenamefont {McClellan}(2019)}]{1cit20}%
  \BibitemOpen
  \bibfield  {author} {\bibinfo {author} {\bibfnamefont {G.~E.}\ \bibnamefont {McClellan}},\ }\bibfield  {title} {\bibinfo {title} {Using raising and lowering operators from geometric algebra for electroweak theory in particle physics},\ }\href@noop {} {\bibfield  {journal} {\bibinfo  {journal} {Adv. Appl. Clifford algebras}\ }\textbf {\bibinfo {volume} {29}} (\bibinfo {year} {2019})}\BibitemShut {NoStop}%
\bibitem [{\citenamefont {McClellan}(2017)}]{1cit21}%
  \BibitemOpen
  \bibfield  {author} {\bibinfo {author} {\bibfnamefont {G.}~\bibnamefont {McClellan}},\ }\bibfield  {title} {\bibinfo {title} {Application of geometric algebra to the electroweak sector of the standard model of particle physics},\ }\href@noop {} {\bibfield  {journal} {\bibinfo  {journal} {Adv. Appl. Clifford algebras}\ }\textbf {\bibinfo {volume} {28}} (\bibinfo {year} {2017})}\BibitemShut {NoStop}%
\bibitem [{\citenamefont {Pavsic}(2004)}]{1cit_1}%
  \BibitemOpen
  \bibfield  {author} {\bibinfo {author} {\bibfnamefont {M.}~\bibnamefont {Pavsic}},\ }\bibfield  {title} {\bibinfo {title} {Kaluza-klein theory without extra dimensions:curved clifford space},\ }\href@noop {} {\bibfield  {journal} {\bibinfo  {journal} {Phys. Lett. B}\ }\textbf {\bibinfo {volume} {614}} (\bibinfo {year} {2004})}\BibitemShut {NoStop}%
\bibitem [{\citenamefont {Pavsic}(2005)}]{1cit_2}%
  \BibitemOpen
  \bibfield  {author} {\bibinfo {author} {\bibfnamefont {M.}~\bibnamefont {Pavsic}},\ }\bibfield  {title} {\bibinfo {title} {Spin gauge theory of gravity in clifford space:a realization of kaluza-klein theory in 4-dimensional spacetime},\ }\href@noop {} {\bibfield  {journal} {\bibinfo  {journal} {Int. J. Mod. Physics. A}\ }\textbf {\bibinfo {volume} {21}} (\bibinfo {year} {2005})}\BibitemShut {NoStop}%
\bibitem [{\citenamefont {Pavsic}(2010{\natexlab{a}})}]{1cit_3}%
  \BibitemOpen
  \bibfield  {author} {\bibinfo {author} {\bibfnamefont {M.}~\bibnamefont {Pavsic}},\ }\bibfield  {title} {\bibinfo {title} {Space inversion of spinors revisited: A possible explanation of chiral behavior in weak interactions},\ }\href@noop {} {\bibfield  {journal} {\bibinfo  {journal} {Phys. Lett. B}\ }\textbf {\bibinfo {volume} {692}} (\bibinfo {year} {2010}{\natexlab{a}})}\BibitemShut {NoStop}%
\bibitem [{\citenamefont {Pavsic}(2010{\natexlab{b}})}]{1cit_4}%
  \BibitemOpen
  \bibfield  {author} {\bibinfo {author} {\bibfnamefont {M.}~\bibnamefont {Pavsic}},\ }\bibfield  {title} {\bibinfo {title} {On the unification of interactions by clifford algebra},\ }\href@noop {} {\bibfield  {journal} {\bibinfo  {journal} {Adv. Appl. Clifford algebras}\ }\textbf {\bibinfo {volume} {20}} (\bibinfo {year} {2010}{\natexlab{b}})}\BibitemShut {NoStop}%
\bibitem [{\citenamefont {Lasenby}\ \emph {et~al.}(1993)\citenamefont {Lasenby}, \citenamefont {Doran},\ and\ \citenamefont {Gull}}]{1cit9}%
  \BibitemOpen
  \bibfield  {author} {\bibinfo {author} {\bibfnamefont {A.}~\bibnamefont {Lasenby}}, \bibinfo {author} {\bibfnamefont {C.}~\bibnamefont {Doran}},\ and\ \bibinfo {author} {\bibfnamefont {S.}~\bibnamefont {Gull}},\ }\bibfield  {title} {\bibinfo {title} {A multivector derivative approach to lagrangian field theory},\ }\href@noop {} {\bibfield  {journal} {\bibinfo  {journal} {Found. Phys}\ }\textbf {\bibinfo {volume} {23}} (\bibinfo {year} {1993})}\BibitemShut {NoStop}%
\bibitem [{\citenamefont {Chisholm}\ and\ \citenamefont {Farwell}(1998)}]{1cit11}%
  \BibitemOpen
  \bibfield  {author} {\bibinfo {author} {\bibfnamefont {J.~S.~R.}\ \bibnamefont {Chisholm}}\ and\ \bibinfo {author} {\bibfnamefont {R.~S.}\ \bibnamefont {Farwell}},\ }\bibinfo {title} {Spin gauge theories: A summary},\ in\ \href@noop {} {\emph {\bibinfo {booktitle} {Clifford Algebras and Their Application in Mathematical Physics: Aachen 1996}}}\ (\bibinfo  {publisher} {Springer Netherlands},\ \bibinfo {year} {1998})\ pp.\ \bibinfo {pages} {53--56}\BibitemShut {NoStop}%
\bibitem [{\citenamefont {J.S.R.~Chisholm}(1999)}]{1cit16}%
  \BibitemOpen
  \bibfield  {author} {\bibinfo {author} {\bibfnamefont {R.~F.}\ \bibnamefont {J.S.R.~Chisholm}},\ }\bibfield  {title} {\bibinfo {title} {Gauge transformations of spinor within a clifford algebraic structure},\ }\href@noop {} {\bibfield  {journal} {\bibinfo  {journal} {J. Physics. A}\ }\textbf {\bibinfo {volume} {32}} (\bibinfo {year} {1999})}\BibitemShut {NoStop}%
\bibitem [{\citenamefont {Weinberg}(1995)}]{2cit1}%
  \BibitemOpen
  \bibfield  {author} {\bibinfo {author} {\bibfnamefont {S.}~\bibnamefont {Weinberg}},\ }\href@noop {} {\emph {\bibinfo {title} {Steven Weinberg}}}\ (\bibinfo  {publisher} {Cambridge University Press},\ \bibinfo {year} {1995})\BibitemShut {NoStop}%
\bibitem [{\citenamefont {Peskin}\ and\ \citenamefont {Schroeder}(1995)}]{Peskin:1995ev}%
  \BibitemOpen
  \bibfield  {author} {\bibinfo {author} {\bibfnamefont {M.~E.}\ \bibnamefont {Peskin}}\ and\ \bibinfo {author} {\bibfnamefont {D.~V.}\ \bibnamefont {Schroeder}},\ }\href@noop {} {\emph {\bibinfo {title} {{An Introduction to quantum field theory}}}}\ (\bibinfo  {publisher} {Addison-Wesley},\ \bibinfo {address} {Reading, USA},\ \bibinfo {year} {1995})\BibitemShut {NoStop}%
\bibitem [{\citenamefont {Ablamowicz}(1995)}]{3cit1}%
  \BibitemOpen
  \bibfield  {author} {\bibinfo {author} {\bibfnamefont {R.}~\bibnamefont {Ablamowicz}},\ }\bibfield  {title} {\bibinfo {title} {Construction of spinors via witt decomposition and primitive idempotents: a review},\ }\href@noop {} {\bibfield  {journal} {\bibinfo  {journal} {Clifford Algebras and Spinor Structures. Mathematics and its Applications Springer}\ }\textbf {\bibinfo {volume} {31}} (\bibinfo {year} {1995})}\BibitemShut {NoStop}%
\bibitem [{\citenamefont {Nakahara}(2003)}]{3cit3}%
  \BibitemOpen
  \bibfield  {author} {\bibinfo {author} {\bibfnamefont {M.}~\bibnamefont {Nakahara}},\ }\href@noop {} {\emph {\bibinfo {title} {Geometry, topology and physics}}}\ (\bibinfo  {publisher} {CRC Press},\ \bibinfo {year} {2003})\BibitemShut {NoStop}%
\bibitem [{\citenamefont {Bailin}\ and\ \citenamefont {A.Love}(1987)}]{3cit2}%
  \BibitemOpen
  \bibfield  {author} {\bibinfo {author} {\bibfnamefont {D.}~\bibnamefont {Bailin}}\ and\ \bibinfo {author} {\bibnamefont {A.Love}},\ }\bibfield  {title} {\bibinfo {title} {Kaluza-klein theories},\ }\href@noop {} {\bibfield  {journal} {\bibinfo  {journal} {Rept. Prog. Phys}\ }\textbf {\bibinfo {volume} {50}} (\bibinfo {year} {1987})}\BibitemShut {NoStop}%
\bibitem [{\citenamefont {Doran}(1994)}]{1cit18}%
  \BibitemOpen
  \bibfield  {author} {\bibinfo {author} {\bibfnamefont {C.}~\bibnamefont {Doran}},\ }\emph {\bibinfo {title} {Geometric algebra and tis application to Mathematical Physics}},\ \href@noop {} {Ph.D. thesis},\ \bibinfo  {school} {University of Cambridge} (\bibinfo {year} {1994})\BibitemShut {NoStop}%
\bibitem [{\citenamefont {Srednicki}(2007)}]{4cit1}%
  \BibitemOpen
  \bibfield  {author} {\bibinfo {author} {\bibfnamefont {M.}~\bibnamefont {Srednicki}},\ }\href@noop {} {\emph {\bibinfo {title} {Quantum Field Theory}}}\ (\bibinfo  {publisher} {Cambridge University Press},\ \bibinfo {year} {2007})\BibitemShut {NoStop}%
\bibitem [{\citenamefont {Walker}(2020)}]{4cit2}%
  \BibitemOpen
  \bibfield  {author} {\bibinfo {author} {\bibfnamefont {M.}~\bibnamefont {Walker}},\ }\bibfield  {title} {\bibinfo {title} {$su(2)\times su(2)$ algebras and the lorentz group $o(3,3)$},\ }\href@noop {} {\bibfield  {journal} {\bibinfo  {journal} {Symmetry}\ }\textbf {\bibinfo {volume} {12}} (\bibinfo {year} {2020})}\BibitemShut {NoStop}%
\bibitem [{\citenamefont {Roldao~da Rocha}(2008)}]{4cit4}%
  \BibitemOpen
  \bibfield  {author} {\bibinfo {author} {\bibfnamefont {J.~V.~J.}\ \bibnamefont {Roldao~da Rocha}},\ }\bibfield  {title} {\bibinfo {title} {Revisiting clifford algebras and spinors iii: Conformal structures and twistors in the paravector model of spacetime},\ }\href@noop {} {\bibfield  {journal} {\bibinfo  {journal} {Int.J.Mod.Physics}\ }\textbf {\bibinfo {volume} {4}} (\bibinfo {year} {2008})}\BibitemShut {NoStop}%
\bibitem [{\citenamefont {Roldao~da Rocha}(2007)}]{4cit3}%
  \BibitemOpen
  \bibfield  {author} {\bibinfo {author} {\bibfnamefont {J.~V.~J.}\ \bibnamefont {Roldao~da Rocha}},\ }\bibfield  {title} {\bibinfo {title} {Conformal structures and twistors in the paravector model of spacetime},\ }\href@noop {} {\bibfield  {journal} {\bibinfo  {journal} {Int.J.Mod.Physics}\ }\textbf {\bibinfo {volume} {4}} (\bibinfo {year} {2007})}\BibitemShut {NoStop}%
\end{thebibliography}%
\end{document}